\documentclass[showpacs,preprintnumbers,amsmath,amssymb,prb]{revtex4}

\def\XXint#1#2#3{{\setbox0=\hbox{$#1{#2#3}{\int}$}
     \vcenter{\hbox{$#2#3$}}\kern-.5\wd0}}

\usepackage{graphicx}
\usepackage{subfigure}
\usepackage{bm}

\begin{document}

\title{Low-energy singlet excitations in spin-$\frac12$ Heisenberg antiferromagnet on square lattice}

\author{A.\ Yu.\ Aktersky$^{1,2}$}
\email{aktersky@gmail.com}
\author{A.\ V.\ Syromyatnikov$^{1,3}$}
\email{asyromyatnikov@yandex.ru}
\affiliation{$^1$National Research Center "Kurchatov Institute" B.P.\ Konstantinov Petersburg Nuclear Physics Institute, Gatchina 188300, Russia}
\affiliation{$^2$Saint Petersburg Academic University --- Nanotechnology Research and Education Centre of the Russian Academy of Sciences, Khlopina 8/3, St.\ Petersburg 194021, Russia}
\affiliation{$^3$Department of Physics, Saint Petersburg State University, Ulianovskaya 1, St.\ Petersburg 198504, Russia}

\date{\today}

\begin{abstract}

We present an approach based on a dimer expansion which describes low-energy singlet excitations (singlons) in spin-$\frac12$ Heisenberg antiferromagnet on simple square lattice. An operator (``effective Hamiltonian'') is constructed whose eigenvalues give the singlon spectrum. The ``effective Hamiltonian'' looks like a Hamiltonian of a spin-$\frac12$ magnet in strong external magnetic field and it has a gapped spectrum. It is found that singlet states lie above triplet ones (magnons) in the whole Brillouin zone except in the vicinity of the point $(\pi,0)$, where their energies are slightly smaller. Based on this finding, we suggest that a magnon decay is possible near $(\pi,0)$ into another magnon and a singlon which may contribute to the dip of the magnon spectrum near $(\pi,0)$ and reduce the magnon lifetime. It is pointed out that the singlon-magnon continuum may contribute to the continuum of excitations observed recently near $(\pi,0)$.

\end{abstract}

\pacs{75.10.Jm, 75.10.-b}

\maketitle

\section{Introduction}

Spin-$\frac12$ Heisenberg antiferromagnet (HAF) on square lattice is one of the most extensively discussed models of quantum magnetism. Interest in this model is particularly stimulated by its relevance to the parent compounds of the high temperature cuprate superconductors. \cite{monous} Despite its simplicity and much theoretical and experimental efforts in studying this model and related compounds, it continues to give surprises.

We discuss the simplest variant of this model which Hamiltonian has the form
\begin{equation}
\label{ham}
{\cal H} = \sum_{\langle i,j \rangle}	{\bf S}_i{\bf S}_j,
\end{equation}
where $\langle i,j \rangle$ denote nearest neighbor spins and the exchange coupling constant is taken to be equal to unity. As HAF \eqref{ham} is defined on bipartite lattice, its ground state is a singlet according to the Marshall's theorem. \cite{marsh,marsh2,auer} The singlet nature of the ground state does not conflict with the long-range N\'eel magnetic order obtained both theoretically and experimentally. \cite{monous} Although no direction is selected in the singlet ground state, the spin-spin correlation function is finite at large distances implying a net staggered magnetization. \cite{liang,dagotto} The spontaneous symmetry breaking takes place in the strict thermodynamic limit accompanied by formation of Goldstone elementary excitations (magnons) carrying spin 1. \cite{chak,monous,reger,auer} Semiclassical approaches based on spin-wave theory ($1/S$-expansion) give surprisingly accurate analytical description of the majority of ground-state and low-temperature properties of this quantum model. \cite{monous} In particular, the low-energy part of the magnon spectrum $\varepsilon_{\bf k}$ is successfully described within the first order in $1/S$ whereas contributions from higher-order terms are very small. \cite{hamer,weih,igar,monous,syromyat} 

\begin{figure}
\includegraphics[scale=0.2]{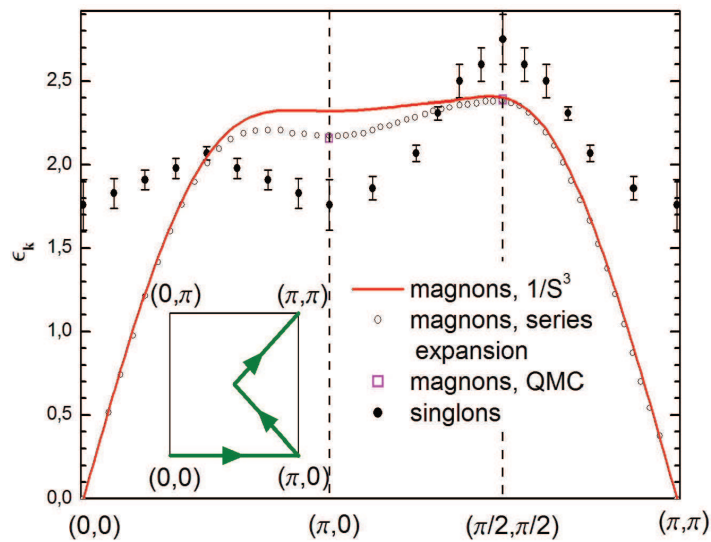}
\caption{(Color online.) Spectrum of spin-$\frac12$ HAF along high-symmetry paths of the Brillouin zone shown in the inset. Spectra of triplet excitations (magnons) are presented obtained in the third order in $1/S$, \cite{syromyat} by series expansion around the Ising limit, \cite{ser} and by Quantum Monte-Carlo (QMC) computation \cite{sand} (available only for ${\bf k}=(\pi,0)$ and $(\pi/2,\pi/2)$). Series expansion and QMC results are consistent with those found recently by continuous similarity transformation technique \cite{uhrig} and they describe quantitatively the magnon spectrum observed experimentally \cite{chris1,piazza} in $\rm Cu(DCOO)_2\cdot4D_2O$. The spectrum is also presented of low-energy singlet excitations (singlons) obtained in the present paper.
\label{spec}}
\end{figure}

The focus of research is currently on short-wavelength magnons. Recent theoretical and experimental results show that $\varepsilon_{\bf k}$ is not reproduced quantitatively by analytical methods around ${\bf k}=(\pi,0)$. A roton-like dip is found in the magnon spectrum at $(\pi,0)$ by Quantum Monte-Carlo, \cite{qmc2,sand} series expansion around the Ising limit, \cite{ser} and continuous similarity transformation \cite{uhrig} techniques (see Fig.~\ref{spec}). Remarkable agreement between these very different numerical methods signifies that the anomaly in the spectrum around $(\pi,0)$ is an intrinsic property of the model. Magnon spectrum extracted from the neutron scattering data obtained in the metal-organic compound ${\rm Cu(DCOO)_2 \cdot 4D_2O}$ (CFTD) which is a perfect realization of model \eqref{ham} shows the roton-like dip at $(\pi,0)$. This dip is quantitatively described by the numerical results mentioned above. \cite{chris1,piazza,cftd} Apart from the magnon peak at energy transfer $\omega\approx\varepsilon_{(\pi,0)}\approx 2.19$, a continuum of excitations is observed extending from the energy slightly below $\varepsilon_{(\pi,0)}$ (from $\omega\approx1.9$) up to $\omega\approx3.8$. \cite{piazza} It is concluded in Ref.~\cite{piazza} that some correlations lead to this continuum which are isotropic in spin space. Continuums of magnetic excitations around $(\pi,0)$ have been observed also experimentally in some cuprates. \cite{cupr1,cupr2,cupr3,cupr4} 
Fractional elementary excitations producing continuums in spin-spin correlators are frequent in quantum physics of one dimension. \cite{1d}
To account for the continuum of excitations, fractional spin-$\frac12$ excitations (spinons) are discussed now in HAF as well. \cite{ho,piazza,tang} It is proposed that a magnon is a confined state of two spinons in the whole Brillouin zone except in the neighborhood of the point $(\pi,0)$, where spinons are deconfined and they form the continuum of excitations. However, it is proposed in Ref.~\cite{uhrig} that all features discussed can be explained in terms of magnons interaction not invoking spinons (while this paper does not disprove spinons existence). 

\begin{figure}
\includegraphics[scale=0.30]{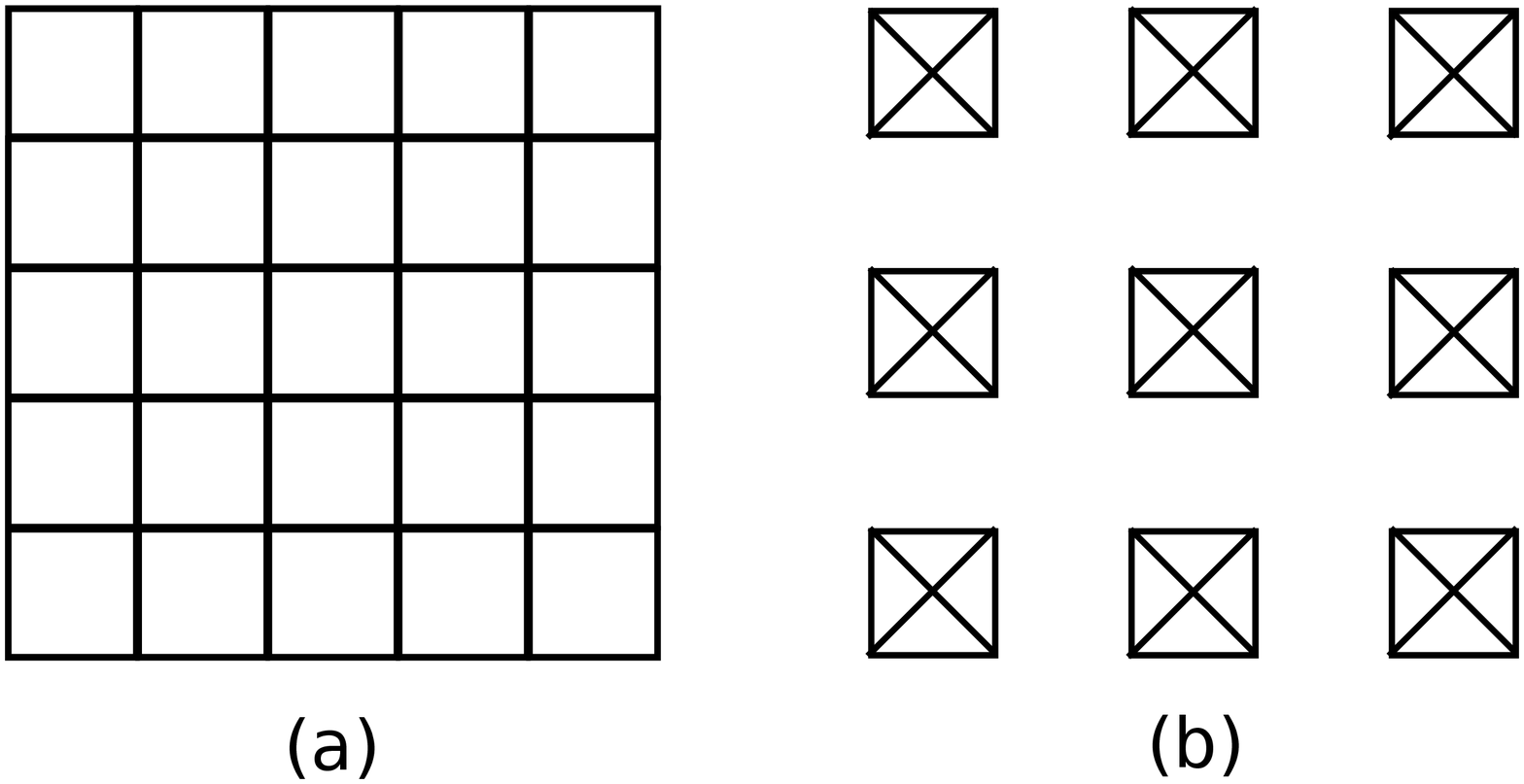}
\caption{
Simple square lattice (a) and decoupled plaquettes having doubly degenerate singlet ground state (b). To come from decoupled plaquettes to the square lattice, we introduce operator \eqref{v} controlled by parameter $\lambda$ so that $\lambda=1$ and $\lambda=0$ correspond to panels (a) and (b), respectively.
\label{lattice}}
\end{figure}

Motivated by the observation of the isotropic excitation continuum which extends to quite large energy, we address the problem of low-energy singlet excitations in model \eqref{ham}. To the best of our knowledge, only triplet excitations are discussed in the literature (spin-$\frac12$ spinons arise as parts of magnons) whereas a little is known about low-energy singlet excitations. We perform a sort of dimer expansion (``plaquette expansion'') to derive an operator (an ``effective Hamiltonian'') whose eigenvalues give the spectrum of singlet excitations (singlons). Our starting point is a set of isolated plaquettes in which exchange coupling constants are equal to unity between all four spins (see Fig.~\ref{lattice}). To come from decoupled plaquettes to the square lattice, we introduce an operator controlled by a single parameter $\lambda$ that switches on interactions between spins from different plaquettes and weakens interactions between diagonal spins in each plaquette. Decoupled plaquettes and model \eqref{ham} correspond to $\lambda=0$ and $\lambda=1$, respectively. Then, we perform perturbation calculations up to 7-th order in $\lambda$ and extrapolate results to $\lambda=1$ by means of standard methods developed for critical phenomena. We present arguments that it is the low-energy singlet sector of HAF \eqref{ham} that we consider in this way.

It is clear that a quantum phase transition (QPT) occurs on the way from decoupled plaquettes to the square lattice because ground state is disordered and magnetically ordered at the beginning and at the end of this way, respectively. The ground state and low-energy excitations change drastically upon QPT that is accompanied by non-analytic behavior of physical quantities as functions of driving parameter near QPT. \cite{sachdev} As the long-range magnetic order cannot arise in the first few orders of perturbation theory, 
\footnote{It is demonstrated in Ref.~\cite{liang} that the ground state wave function in the ordered phase is a superposition of wave functions at least some of which contain products of singlet states of quite distant spin pairs. It is clear that such states can appear in dimer expansions only in high orders in $\lambda$.}
dimer expansions are inappropriate for discussion of ground-state properties in the ordered phase. It is the reason why authors of previous dimer expansions \cite{aff,singh} do not consider the ordered phase focusing on properties of QPTs from disordered phases to the ordered one. However singlet excitations lie well above long-wavelength triplet ones (see below) which determine properties of the QPT. As high-energy excitations normally do not change drastically upon phase transitions, this gives promise that dimer expansions can describe singlet excitations. That is why we develop below the special dimer expansion which is convenient for discussion of the low-energy singlet excitations. This expansion differs from those suggested in Refs.~\cite{aff,singh} in which starting points are isolated couples of spins.

The main result of our study (i.e., the singlon spectrum) is illustrated by Fig.~\ref{spec}. We find that the ``effective Hamiltonian'' looks like a Hamiltonian of a spin-$\frac12$ magnet in strong external magnetic field and that the singlon spectrum is gapped (the large gap value is in agreement with previous findings according to which the low-energy physics of HAF is governed by low-energy magnons). Singlet states lie above triplet ones (magnons) in the whole Brillouin zone except in the vicinity of the point $(\pi,0)$. Remarkably, the singlon spectrum crosses the magnon one at those points, where analytical results of the $1/S$-expansion start to deviate from numerical data for the magnon spectrum. Based on this finding, we suggest that a magnon decay is possible near $(\pi,0)$ into another magnon and a singlon which may contribute to the dip of the magnon spectrum near $(\pi,0)$ and reduce the magnon lifetime. Notice that such a microscopic mechanism fundamentally cannot arise in $1/S$-expansion. We point out that one-singlon states are invisible for experimental methods measuring two-spin correlators. However triplet states containing one magnon and one singlon and forming a singlon-magnon continuum may contribute to the experimentally observed continuum of excitations near $(\pi,0)$. It should be stressed that our findings and conclusions cannot disprove neither of the physical pictures proposed before for the description of the high-energy peculiarities of model \eqref{ham}. In the present stage, we can only point out the possibility of additional contributions from singlons to the discussed features.

The rest of the present paper is organized as follows. We describe our approach in Section~\ref{plaqexp}. The ``effective Hamiltonian'' is constructed and analyzed in Sections~\ref{hamcon} and \ref{haman}, respectively. Results obtained are discussed in Section~\ref{resdis}. Section~\ref{conc} contains our conclusion.

\section{Plaquette expansion}
\label{plaqexp}

\begin{figure}
\includegraphics[scale=0.40]{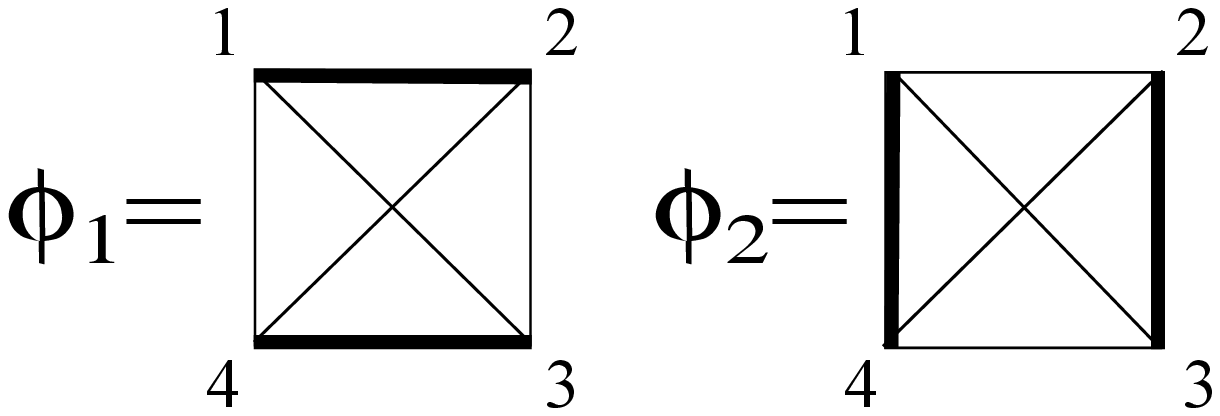}
\caption{ 
Wave functions of an isolated plaquette from which its singlet ground state wave functions \eqref{gswf} are constructed. Bold lines denote singlet states of the corresponding two spins.
\label{bonds}}
\end{figure}

Wave functions of the doubly degenerate singlet ground state of isolated plaquette
\begin{equation}
\label{gswf}
\begin{aligned}
\Psi^+ &= \frac{\phi_1+\phi_2}{\sqrt{3}},\\
\Psi^- &= \phi_1-\phi_2
\end{aligned}
\end{equation}
are constructed as linear combinations of nonorthogonal ones
$
\phi_1=(|\uparrow\rangle_1|\downarrow\rangle_2 
- 
|\downarrow\rangle_1|\uparrow\rangle_2)
(|\uparrow\rangle_3|\downarrow\rangle_4
-
|\downarrow\rangle_3|\uparrow\rangle_4)/2
$
and
$
\phi_2=(|\uparrow\rangle_2|\downarrow\rangle_3 
- 
|\downarrow\rangle_2|\uparrow\rangle_3)
(|\uparrow\rangle_4|\downarrow\rangle_1
-
|\downarrow\rangle_4|\uparrow\rangle_1)/2
$
which are depicted in Fig.~\ref{bonds}. It is convenient to consider $\Psi^+$ and $\Psi^-$ as states of a pseudospin $s=\frac12$ corresponding to $s_z=1/2$ and $s_z=-1/2$, respectively.

To come from decoupled plaquettes to the square lattice, one has to switch on the following spin interactions which magnitudes are controlled by the single parameter $\lambda$:
\begin{equation}
\label{v}
V = 
\lambda 
\left( \sum_{\langle i,j\rangle} 
\left({\bf S}^{(i)}_1{\bf S}^{(j)}_2+{\bf S}^{(i)}_3{\bf S}^{(j)}_4\right)
+
\sum_{\langle i,p\rangle} 
\left({\bf S}^{(i)}_1{\bf S}^{(p)}_4+{\bf S}^{(i)}_2{\bf S}^{(p)}_3\right)
-
\sum_i \left({\bf S}^{(i)}_1{\bf S}^{(i)}_3+{\bf S}^{(i)}_2{\bf S}^{(i)}_4\right)
\right),
\end{equation}
where upper and lower indexes of $\bf S$ enumerate plaquettes and the spin number in the plaquette (according to Fig.~\ref{bonds}), respectively, and $\langle i,j\rangle$ and $\langle i,p\rangle$ denote nearest neighbor plaquettes in horizontal and vertical directions, correspondingly. Decoupled plaquettes and HAF on the square lattice \eqref{ham} correspond to $\lambda=0$ and $\lambda=1$, respectively. 

The singlet ground state is $2^N$ times degenerate in the system containing $N$ decoupled plaquettes. This degenerate energy level splits into a singlet band (at $N\to\infty$) upon $\lambda$ increasing because $V$ commutes with operator of the total spin. Our goal is to describe this band in terms of interaction between pseudospins. It is our hope that at least lower part of this band forms the low-energy singlet sector of HAF \eqref{ham}. It may happen, however, that some upper singlet levels cross all energy levels stemming from the lowest singlet one upon $\lambda$ increasing from 0 to 1. In this case, all the band of singlet excitations we consider below would describe some excited states of HAF. We suggest the following symmetry arguments against this scenario. It is important that $\Psi^+$ and $\Psi^-$ given by Eq.~\eqref{gswf} belong to different irreducible representations of the point group $C_{4v}$ of the plaquette. Then, in any cluster having the symmetry $C_{4v}$ and containing $N$ plaquettes (see Fig.~\ref{lattice}(b) for the smallest cluster of this kind), one can find functions belonging to any irreducible representation of the group $C_{4v}$ among $2^N$ ground state wave functions. If the perturbation is one-parametric, the level crossing is generally forbidden of two energy levels belonging to the same irreducible representation. \cite{gth} Then, the lower singlet levels stemming from the lower degenerate level of such clusters would repel all the upper singlet levels upon $\lambda$ increasing from 0 to 1. Bearing in mind that one can consider arbitrary large clusters of this kind and that boundary conditions in a large enough system have normally a very small impact on its bulk properties, it is reasonable to conclude that we deal with the singlet band at least lower part of which forms the lowest singlet sector of HAF \eqref{ham}.

\section{``Effective Hamiltonian'' construction}
\label{hamcon}

Our main goal is to obtain the operator $H$ (the ``effective Hamiltonian'') acting in the pseudospin space which eigenvalues give energies of levels in the lower singlet band (assuming that $N\to\infty$). Consideration of the corresponding wave functions in the spin space is out of the scope of the present paper. We use the standard perturbation theory for systems with degenerate energy levels developed by Bloch and described, e.g., in the textbook \cite{messi}. The conventional series expansion technique has a similar background. \cite{gel} 

To illustrate the idea, let us consider corrections of orders $\lambda^0$ and $\lambda^1$. In the zeroth order, $H=CN$, where $C=-\lambda^03/2$ is given by the ground state energy of isolated plaquette. The first order corrections to energies of lower singlet levels are determined by matrix elements 
$\langle \Psi^\pm_1\Psi^\pm_2\dots\Psi^\pm_N 
|V| 
\Psi^\pm_1\Psi^\pm_2\dots\Psi^\pm_N\rangle$, 
where $V$ is given by Eq.~\eqref{v} and $\Psi^\pm_i$ denote states \eqref{gswf} of $i$-th plaquette. It is easy to check that the first two terms in Eq.~\eqref{v} transform a ground state wave function of a plaquette to the triplet sector. As a result nonzero corrections in the first order in $\lambda$ originate only from the third term in Eq.~\eqref{v} which contains interactions between spins belonging to the same plaquette. In terms of pseudospins, these corrections can be described by an interaction of each pseudospin with an effective magnetic field $h$: 
$ H = CN+ h\sum_i s_i^z $,
where 
$ C = -\lambda^03/2 + \frac12
(\langle \Psi^-\Psi^- |V_3| \Psi^-\Psi^- \rangle
+
\langle \Psi^+\Psi^+ |V_3| \Psi^+\Psi^+ \rangle
) = -\lambda^03/2+\lambda/2 $,
$ h = \langle \Psi^-\Psi^- |V_3| \Psi^-\Psi^- \rangle
-
\langle \Psi^+\Psi^+ |V_3| \Psi^+\Psi^+ \rangle
=2\lambda $,
and $V_3 = -\lambda({\bf S}_1{\bf S}_3+{\bf S}_2{\bf S}_4)$ (cf.\ the third term in Eq.~\eqref{v}). 

Terms of the perturbation theory of the second order in $\lambda$ contribute to $C$, $h$, and they lead also to terms describing an interaction between nearest-neighbor pseudospins. As a result $H$ acquires the form in the second order in $\lambda$
\begin{eqnarray}
\label{efh}
H &=& CN + h\sum_i s_i^z
+ \sum_{\langle i,j\rangle} 
\left(J^{zz}_{ij}s_i^z s_j^z
+ J^{+z}_{ij}s_i^+ s_j^z
+ J^{-z}_{ij}s_i^- s_j^z
+ J^{z+}_{ij}s_i^z s_j^+
+ J^{z-}_{ij}s_i^z s_j^-
+ J^{++}_{ij}s_i^+ s_j^+
\right.\nonumber\\
&&{}\left.
+ J^{--}_{ij}s_i^- s_j^-
+ J^{-+}_{ij}s_i^- s_j^+
+ J^{+-}_{ij}s_i^+ s_j^-\right).
\end{eqnarray}
Apart from corrections to coefficients in Eq.~\eqref{efh}, higher-order terms of the perturbation theory lead also to multi-pseudospin interactions and to two-pseudospin long-range interactions.

\begin{table}
\caption{
Nonzero coefficients of the ``effective Hamiltonian'' $H$ in the first seven orders in $\lambda$ describing interaction between nearest and next-nearest neighbor pseudospins. Numerous coefficients for long-range and multi-pseudospin interactions have been also calculated but they are not presented here. Subscripts $h$, $v$, and $d$ denote the shortest horizontal, vertical, and diagonal bonds, respectively. The constant $C$ has also the zero-order correction equal to $-\lambda^0 3/2$ which is given by the ground state energy of an isolated plaquette.
\label{coef}
}
\begin{ruledtabular}
\begin{tabular}{cccccccc}
& 1&2&3&4&5&6&7\\
\hline
$C$ &  0.5 & -0.416666 & -0.250000 & -0.266783 & -0.315197 & -0.486108 & -0.807964\\
$h$ & 2.0 & 0.166667 & -0.375000 & -0.574073 & -1.031603 & -1.746714 & -3.619268\\
\hline
$J^{zz}_h$ &0 & -0.083333 & -0.062500 & -0.218171 & -0.436133 & -0.937999 & -2.063095\\
$J^{--}_h$ &0 & -0.062500 & 0.015625 & -0.031684 & 0.031262 & -0.138602 & -0.195048\\
$J^{++}_h$ &0 & -0.062500 & -0.109375 & -0.219184 & -0.463558 & -1.208478 & -2.851198\\
$J^{+-}_h$ &0 & -0.062500 & -0.046875 & -0.062934 & -0.058693 & -0.071379 & -0.063827\\
$J^{-+}_h$ &0 & -0.062500 & -0.046875 & -0.062934 & -0.058693 & -0.071379 & -0.063827\\
$J^{-z}_h$ &0 & -0.072169 & -0.018042 & -0.002506 & -0.027965 & -0.051093 & -0.238453\\
$J^{+z}_h$ &0 & -0.072169 & -0.090211 & -0.110759 & -0.143319 & -0.206620 & -0.342623\\
$J^{z-}_h$ &0 & -0.072169 & -0.018042 & -0.002506 & -0.027965 & -0.051093 & -0.238453\\
$J^{z+}_h$ &0 & -0.072169 & -0.090211 & -0.110759 & -0.143319 & -0.206620 & -0.342623\\
\hline
$J_d^{zz}$ &0 & 0 & 0 & 0.012732 & 0.054546 & 0.010743 & -0.177531\\
$J_d^{--}$ &0 & 0 & 0 & -0.009549 & 0.002464 & -0.026063 & 0.023463\\
$J_d^{++}$ &0 & 0 & 0 & -0.009549 & -0.040996 & -0.157527 & -0.502310\\
$J_d^{-+}$ &0 & 0 & 0 & -0.009549 & -0.019266 & -0.038716 & -0.058577\\
$J_d^{+-}$ &0 & 0 & 0 & -0.009549 & -0.019266 & -0.038716 & -0.058577\\
\hline
$J_v^{zz}$ &0 & -0.083333 & -0.062500 & -0.218171 & -0.436133 & -0.937999 & -2.063095\\
$J_v^{--}$ &0 & -0.062500 & 0.015625 & -0.031684 & 0.031262 & -0.138602 & -0.195048\\
$J_v^{++}$ &0 & -0.062500 & -0.109375 & -0.219184 & -0.463558 & -1.208478 & -2.851198\\
$J_v^{+-}$ &0 & -0.062500 & -0.046875 & -0.062934 & -0.058693 & -0.071379 & -0.063827\\
$J_v^{-+}$ &0 & -0.062500 & -0.046875 & -0.062934 & -0.058693 & -0.071379 & -0.063827\\
$J_v^{-z}$ &0 & 0.072169 & 0.018042 & 0.002506 & 0.027965 & 0.051093 & 0.238453\\
$J_v^{+z}$ &0 & 0.072169 & 0.090211 & 0.110759 & 0.143319 & 0.206620 & 0.342623\\
$J_v^{z-}$ &0 & 0.072169 & 0.018042 & 0.002506 & 0.027965 & 0.051093 & 0.238453\\
$J_v^{z+}$ &0 & 0.072169 & 0.090211 & 0.110759 & 0.143319 & 0.206620 & 0.342623\\
\hline
\end{tabular}
\end{ruledtabular}
\end{table}

We calculate corrections to all parameters in the ``effective Hamiltonian'' $H$ (including the multi-pseudospin and the long-range terms) up to the 7-th order in $\lambda$. Series are presented in Table~\ref{coef} for all the nonzero coefficients excluding numerous coefficients for long-range and multi-pseudospin interactions. It should be noted that in accordance with the general property of the perturbation series, \cite{messi} operator $H$ is non-Hermitian (see, e.g., series for $J^{++}$ and $J^{--}$). We find below that the spectrum of $H$ is real as it must be. Operator $H$ is translationally invariant: it is defined on the square lattice which period twice as large as the period of the original lattice. 
\footnote{
We call $H$ ``effective Hamiltonian'' because it is non-Hermitian and it describes only singlet sector of the initial model (1).
}

To extrapolate the spectrum from $\lambda\ll1$ to $\lambda=1$, we use Pad\'e and Pad\'e-Borel resummation techniques. \cite{pade,gutt,padeBorel} There is no point in applying these techniques individually to each coefficient in $H$ because the number of terms is small in the series for some of them (see, e.g., Table~\ref{coef}). Then, we derive below analytical expressions for physical quantities and find series for them up to the 7-th order in $\lambda$ using series for the coefficients in $H$.

\section{``Effective Hamiltonian'' analysis}
\label{haman}

The resummation procedure gives for $h$ quite a large value around 2 at $\lambda\le1$. Then, it is reasonable to suppose that $\langle s_i^z \rangle \approx -1/2$ in the ``ground state'' of $H$ at $\lambda\le1$ (this assumption is confirmed by the spectrum stability in further calculations). Then, it is convenient to use the conventional Holstein-Primakoff transformation to find the spectrum
\begin{equation}
\label{hp}
s^-_i = \sqrt{2s - a_i^\dagger a_i }\, a_i, \quad
s^+_i = a^\dagger_i \sqrt{2s - a_i^\dagger a_i },\quad
s^z_i = -s + a_i^\dagger a_i.
\end{equation}
Substituting Eqs.~\eqref{hp} into $H$, one leads to the Bose-analog of the ``effective Hamiltonian'' $H = E_0 + \sum_{n=1}^\infty H_n$, where $E_0$ is the ground state energy not renormalized by pseudospin fluctuations and $H_n$ denote terms containing products of $n$ operators $a$ and $a^\dagger$. 

At least the first seven terms in the series for $H_1$ are equal to zero. It is a consequence of the symmetry of series for coefficients contributing to $H_1$. This property can be illustrated by series for $J^{\pm z}$ and $J^{z \pm}$ presented in Table~\ref{coef}. It is seen that series for these coefficients for vertical and horizontal bonds have opposite signs in all orders in $\lambda$. One concludes also from this fact that quantum corrections to $H_1$ from terms $H_n$ with odd $n$ vanish as well.

\begin{figure}
\includegraphics[scale=0.4]{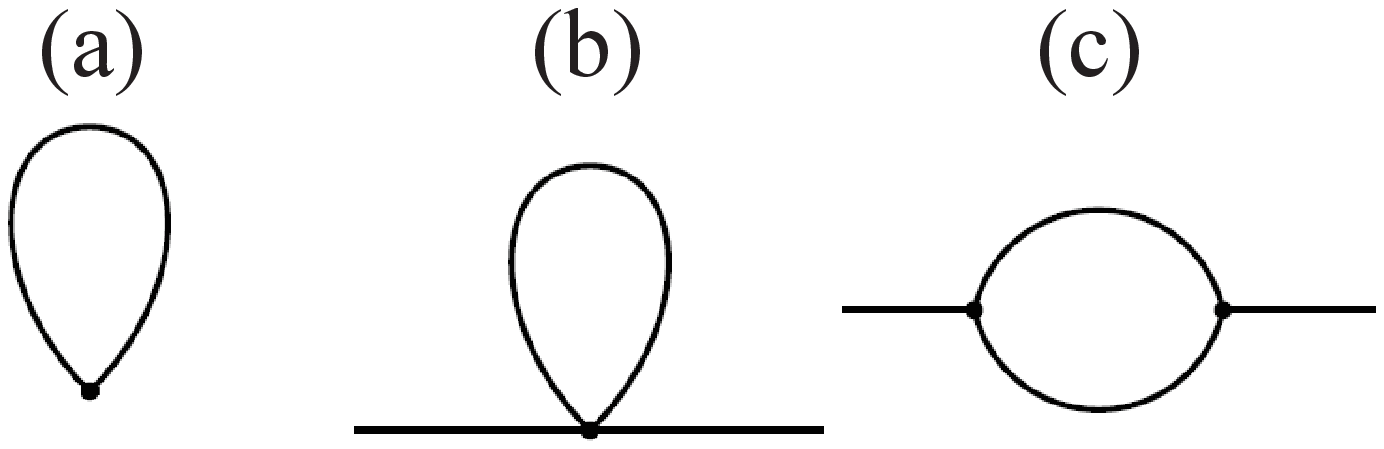}
\caption{
Simplest diagrams contributing to renormalization of the ground state energy (a) and to the singlon spectrum (b)--(c) due to fluctuations of pseudospins.
\label{diagrams}}
\end{figure}

The bilinear part of $H$ has the form
\begin{eqnarray}
\label{h2}
H_2 &=& \sum_{\bf k}
\left(
E_{\bf k} a^\dagger_{\bf k}a_{\bf k}
+
\frac{B_{m\bf k}}{2} a_{\bf k}a_{-\bf k} + \frac{B_{p\bf k}}{2}a^\dagger_{\bf k}a^\dagger_{-\bf k}
\right),
\end{eqnarray}
where $B_{m\bf k} \ne B_{p\bf k}^*$ so that $H_2$ is non-Hermitian. The Bose-analog of $H$ can be analyzed in a standard way (see, e.g., Ref.~\cite{syromyat}) by introducing Green's functions $G(k) = \langle a_{\bf k}, a^\dagger_{\bf k} \rangle_\omega$, $F(k) = \langle a_{\bf k}, a_{-\bf k} \rangle_\omega$, ${\overline G}(k) = \langle a^\dagger_{-\bf k}, a_{-\bf k} \rangle_\omega$ and $F^\dagger (k) = \langle a^\dagger_{-\bf k}, a^\dagger_{\bf k} \rangle_\omega$, where $k=(\omega,{\bf k})$. We have two sets of Dyson equations for them one of which has the form
\begin{equation}
\label{eqfunc}
\begin{array}{l}
G(k) = G^{(0)}(k) + G^{(0)}(k){\overline \Sigma}(k)G(k) + G^{(0)}(k) [B_{p\bf k} + \Pi(k)] F^\dagger(k),\\
F^\dagger(k) = {\overline G}^{(0)}(k) \Sigma(k)F^\dagger(k) + {\overline G}^{(0)}(k) [B_{m\bf k} + \Pi^\dagger(k) ]G(k),
\end{array}
\end{equation}
where $G^{(0)}(k) = (\omega - E_{\bf k})^{-1}$ is the bare Green's function and $\Sigma(k)$, $\overline \Sigma(k)$, $\Pi(k)$, and $\Pi^\dagger(k)$ are self-energy parts. One obtains solving Eqs.~(\ref{eqfunc}) and a similar set of equations for $\overline G(k)$ and $F(k)$
\begin{equation}
\label{gf}
\begin{aligned}
G(k) &= \frac{\omega + E_{\bf k} + \Sigma(k)}{{\cal D}(k)},\quad
\overline G(k) = \frac{-\omega + E_{\bf k} + \overline \Sigma(k)}{{\cal D}(k)},\\
F(k) &= -\frac{B_{p\bf k} + \Pi(k)}{{\cal D}(k)},\quad
F^\dagger(k) = -\frac{B_{m\bf k} + \Pi^\dagger(k)}{{\cal D}(k)},
\end{aligned}
\end{equation}
where
\begin{eqnarray}
\label{d}
{\cal D}(k) &=& \omega^2 - \epsilon_{0\bf k}^2 - \Omega(k),\\
\label{spec0}
\epsilon_{0\bf k} &=& \sqrt{E_{\bf k}^2 - B_{m\bf k}B_{p\bf k}},\\
\label{o}
\Omega(k) &=& E_{\bf k}(\Sigma + \overline{\Sigma}) - B_{m\bf k}\Pi - B_{p\bf k}\Pi^\dagger - \omega (\Sigma - \overline{\Sigma}) - \Pi\Pi^\dagger + \Sigma \overline{\Sigma},
\end{eqnarray}
$G(k)=\overline G(-k)$, $\Sigma(k)=\overline \Sigma(-k)$, and $\epsilon_{0\bf k}$ is the spectrum in the linear spin-wave approximation. Quantity $\Omega(k)$ given by Eq.~(\ref{o}) describes renormalization of the spectrum square. We find $\Omega(k)$ within the first order in $1/s$ by calculating corresponding diagrams for self-energy parts shown in Figs.~\ref{diagrams}(b) and \ref{diagrams}(c). The diagram presented in Fig.~\ref{diagrams}(a) gives the first $1/s$ correction to the ground state energy. All the diagrams give small contributions to the physical quantities due to large gap in the bare spectrum $\epsilon_{0\bf k}$ (see below). They are calculated as follows. We carry out integration over internal energy, expand the resultant expression into a series up to the 7-th order in $\lambda$, and carry out the summation over internal momenta. The external energy is taken to be equal to $\epsilon_{0\bf k}$ in the diagram shown in Fig.~\ref{diagrams}(c). Due to the smallness of $1/s$-corrections, $\lambda$ is actually the only parameter in our theory controlling series for physical quantities.

\section{Results and discussion}
\label{resdis}

{\it Ground state energy.} The series for the ground state energy per spin $E_{gs}$ not renormalized by pseudospin fluctuations is obtained from $H$ by replacing $s_i^z$ and $s_i^\pm$ by $-1/2$ and 0, respectively. Then, the constant term and terms containing only $z$-components of pseudospins contribute to $E_{gs}$.
Correction from the diagram shown in Fig.~\ref{diagrams}(a) leads to only slight variation of coefficients in the series (that does not practically change the result of resummation) which has the form 
$
-0.375-0.125 \lambda-0.135417 \lambda^2-0.0239258 \lambda^3-0.0262953 \lambda^4-0.00190434 \lambda^5-0.0144457 \lambda^6+0.0155128 \lambda^7
$ 
(all coefficients in all series are calculated with machine precision). 
As coefficients in this series do not rise and have irregular signs, we analyze only Pad\'e approximants and obtain
$
E_{gs} = -0.694(5).
$
The estimated errors in all the extrapolated values are always somewhat subjective. \cite{gutt} We exclude first all the approximants with obvious ``defects'' containing spurious poles with low residue at small $\lambda$. The remaining approximants are weighted to favor the higher-order ones. The error is estimated from the scatter among the weighted values.

Previous numerical studies \cite{uhrig,monous,richter1,richter2} give values for the ground state energy around $-0.669$ that differs slightly from our finding. We attribute this discrepancy to the above discussed inability of dimer expansions to describe the ground-state properties in ordered phases. As the long-range magnetic order cannot arise in the first few orders of the perturbation theory, one has to analyze the whole series in $\lambda$ to find, e.g., the ground state energy at $\lambda=1$. Other dimer expansions underestimate $E_{gs}$ as well in the ordered phase of HAF. \cite{aff}

It should be pointed out also that contributions of long-range and multi-pseudospin interactions are not negligible: the value of $E_{gs}$ found using ``effective Hamiltonian'' \eqref{efh} with coefficients from Table~\ref{coef} is $-0.685(5)$.

{\it Spectrum of singlet excitations.} The spectrum is presented in Fig.~\ref{spec} found from the bilinear part of the ``effective Hamiltonian'' \eqref{h2} (i.e., using Eq.~\eqref{spec0}) by means of Pad\'e approximants analysis as well as Pad\'e-Borel resummation approach. In particular, series for the spectrum at ${\bf k}=(\pi,0)$ and $(\pi/2,\pi/2)$ have the form
\begin{eqnarray}
\epsilon_{(\pi,0)} &=& 2 \lambda+0.083334 \lambda^2-0.453125 \lambda^3-0.356698 \lambda^4-0.47287 \lambda^5-0.342342 \lambda^6-0.468077 \lambda^7,\\
\epsilon_{(\pi/2,\pi/2)} &=& 2 \lambda+0.583334 \lambda^2-0.078125 \lambda^3+0.254847 \lambda^4+0.0564162 \lambda^5+0.253933 \lambda^6-0.173274 \lambda^7
\end{eqnarray}
which give after the resummation $1.76\pm0.15$ and $2.75\pm0.15$, respectively. Again, contributions of long-range and multi-pseudospin interactions to these results are not negligible being of the order of 10\%. As the gap in the singlet spectrum is large, one expects a small spectrum renormalization by pseudospin fluctuations. For instance, diagrams shown in Figs.~\ref{diagrams}(b) and \ref{diagrams}(c) do not practically change the results.

It is interesting to speculate about the role of singlet excitations observed above using the quasiparticle concept thus assuming that each excited singlet state corresponds to a quasiparticle (singlon) carrying spin 0. It should be noted that one-singlon states are invisible for experimental methods measuring two-spin correlators. For instance, they give zero contribution to the dynamic structure factor $\propto {\rm Im}\langle S^\alpha_{\bf k}, S^\beta_{-\bf k}\rangle_\omega$ because any operator $S^\alpha_i$ moves the singlet ground state wave function to the triplet sector. 
%\footnote{However, one-singlon states can produce sharp anomalies in the dynamic structure factor (as magnons do) if some (small) interaction arises in the Hamiltonian not conserving the total spin.}
However triplet excited states containing one magnon and one singlon can contribute to two-spin correlators. Such states can form a continuum at any given momentum $\bf k$ determined by the sum $\varepsilon_{\bf q}+\epsilon_{\bf k-q}$ which starts at the singlon energy $\epsilon_{\bf k}$ and ends at $\varepsilon_{\bf k}+\epsilon_{\bf 0}$. Using our results for $\epsilon_{\bf k}$ and those \cite{piazza} for the magnon spectrum $\varepsilon_{\bf k}$, we obtain at ${\bf k}=(\pi,0)$ that this continuum lie within the energy interval $(1.76\pm0.15,3.95\pm0.17)$ which is in agreement with that $(\sim1.9,\sim3.8)$ observed in CFTD experimentally. \cite{piazza} A weak continuum is seen also in the experimental data at $(\pi/2,\pi/2)$ lying in the range $(\sim2.3,\sim3.8)$. Our results give the interval $(2.75\pm0.15,4.14\pm0.17)$ in this case. Thus, the singlon-magnon continuum can contribute to that obtained experimentally.

\begin{figure}
\includegraphics[scale=0.22]{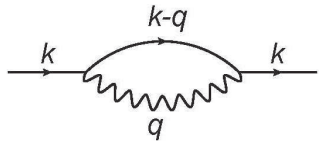}
\caption{
Illustration of the possible microscopic mechanism contributing to the dip in the magnon spectrum near ${\bf k}=(\pi,0)$: magnon decay into another magnon and a singlon. Solid and wavy lines stand for magnon and singlon Green's functions, respectively.
\label{decay}}
\end{figure}

Curiously, the singlon spectrum shown in Fig.~\ref{spec} lies below the magnon one at that region of the Brillouin zone, where analytical results of the $1/S$-expansion deviate from numerical data for the magnon spectrum. Then, one can speculate about the microscopic mechanism contributing to the dip in the magnon spectrum near $(\pi,0)$. As the magnon spectrum enters into the singlon-magnon continuum at that region, a magnon decay may become possible there into another magnon and a singlon (see Fig.~\ref{decay}). In many cases, such decay processes lower the quasiparticle energy and reduce its lifetime. Notice also that such a microscopic mechanism fundamentally cannot arise in $1/S$-expansion. Derivation of an effective Hamiltonian describing both singlet and triplet excitations of the initial model \eqref{ham} would be an interesting subject of further discussion. Such effective Hamiltonian would provide a quantitative description of the processes shown in Fig.~\ref{decay}.

To conclude, singlet excitations may contribute to the high-energy peculiarities of HAF \eqref{ham} and further discussion is required to fully clarify their role in this model.

\section{Conclusion}
\label{conc}

We develop the approach based on the plaquette expansion to discuss the spectrum of lower singlet excitations in spin-$\frac12$ Heisenberg antiferromagnet \eqref{ham} on simple square lattice. The operator is found in the first seven orders of the perturbation expansion whose eigenvalues give the spectrum of low-lying singlet excitations (singlons). The resummation procedure gives the singlon spectrum shown in Fig.~\ref{spec}. We obtain that lower singlet excitations lie below triplet ones near the point ${\bf k}=(\pi,0)$. We suggest that the magnon decay is possible near $(\pi,0)$ into another magnon and a singlon (see Fig.~\ref{decay}) which may contribute to the dip of the magnon spectrum near $(\pi,0)$ and reduce the magnon lifetime. This microscopic mechanism fundamentally cannot arise in $1/S$-expansion. It is pointed out that one-singlon states are invisible for experimental methods measuring two-spin correlators. However the singlon-magnon continuum which can arise in this model may contribute to the continuum of excitations observed recently in CFTD. \cite{piazza}

\begin{acknowledgments}

We thank V.Yu.~Petrov, A.I.~Sokolov, and G.S.~Uhrig for discussion. This work is supported by Russian Scientific Fund Grant No.\ 14-22-00281.

\end{acknowledgments}

\bibliography{bib}

\end{document}